\documentclass[aps,prd,noeprint,twocolumn,showpacs,amsmath,amssymb]{revtex4-2}
\usepackage{amsmath}
\usepackage{graphicx}
\usepackage{subfigure}
\usepackage{epstopdf}
\usepackage{color}
\usepackage{multirow}
\usepackage{setspace}
\usepackage{overpic}
\usepackage{amssymb}
\usepackage[bookmarksnumbered, pdfstartview=FitH,colorlinks,urlcolor=blue, citecolor=blue,linkcolor=blue]{hyperref}
\usepackage{lineno}
\usepackage{bm}
\usepackage{rotating}
\usepackage[utf8]{inputenc}
\newcommand{\xisemi}{\Xi^{0}\to\Sigma^{-}e^{+}\nu_{e}}
\newcommand{\dsdq}{\Delta S=\Delta Q}

\let\oldequation\equation
\let\oldendequation\endequation

\renewenvironment{equation}
  {\linenomathNonumbers\oldequation}
  {\oldendequation\endlinenomath}

\begin{document}

\title{\bf \boldmath
Search for hyperon \texorpdfstring{$\dsdq$}{DeltaS = DeltaQ} violating decay \texorpdfstring{$\xisemi$}{Xi0 to Sigma- e+ nu}}

\author{
\begin{small}
\begin{center}
M.~Ablikim$^{1}$, M.~N.~Achasov$^{11,b}$, P.~Adlarson$^{70}$, M.~Albrecht$^{4}$, R.~Aliberti$^{31}$, A.~Amoroso$^{69A,69C}$, M.~R.~An$^{35}$, Q.~An$^{66,53}$, X.~H.~Bai$^{61}$, Y.~Bai$^{52}$
% ...
\\
\vspace{0.2cm}
(BESIII Collaboration)\\
\vspace{0.2cm} {\it
$^{1}$ Institute of High Energy Physics, Beijing 100049, People's Republic of China\\
% ...
$^{4}$ Bochum Ruhr-University, D-44780 Bochum, Germany\\
% ...
$^{11}$ G.I. Budker Institute of Nuclear Physics SB RAS (BINP), Novosibirsk 630090, Russia\\
% ...
$^{31}$ Johannes Gutenberg University of Mainz, Johann-Joachim-Becher-Weg 45, D-55099 Mainz, Germany\\
% ...
$^{35}$ Liaoning Normal University, Dalian 116029, People's Republic of China\\
% ...
$^{52}$ Southeast University, Nanjing 211100, People's Republic of China\\
$^{53}$ State Key Laboratory of Particle Detection and Electronics, Beijing 100049, Hefei 230026, People's Republic of China\\
% ...
$^{61}$ University of Jinan, Jinan 250022, People's Republic of China\\
% ...
$^{66}$ University of Science and Technology of China, Hefei 230026, People's Republic of China\\
% ...
$^{69}$ University of Turin and INFN, (A)University of Turin, I-10125, Turin, Italy; (B)University of Eastern Piedmont, I-15121, Alessandria, Italy; (C)INFN, I-10125, Turin, Italy\\
$^{70}$ Uppsala University, Box 516, SE-75120 Uppsala, Sweden\\
% ...
$^{b}$ Also at the Novosibirsk State University, Novosibirsk, 630090, Russia\\
% ...
}\end{center}
\end{small}
}

\begin{abstract}
Using a data sample of $(1.0087\pm0.0044)\times10^{10}$ $J/\psi$ decay events collected with the BESIII detector at the center-of-mass energy of $\sqrt{s}=3.097$ GeV, we present a search for the hyperon semileptonic decay $\xisemi$ which violates the $\dsdq$ rule. No significant signal is observed, and the upper limit on the branching fraction $\mathcal{B}(\xisemi)$ is determined to be $1.6\times10^{-4}$ at the 90\% confidence level. This result improves the previous upper limit result by about one order of magnitude.
\end{abstract}


\maketitle

\oddsidemargin  -0.2cm
\evensidemargin -0.2cm

\section{Introduction}
\begin{figure}[htbp]\centering
\includegraphics[width=0.7\linewidth]{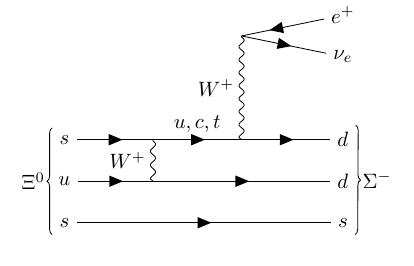}
\caption{The Feynman diagram of the $\dsdq$ violating decay $\xisemi$.}
\label{fig:feynman}
\end{figure}

Hyperon semileptonic decays play an important role in understanding the interplay between weak and strong interactions, where the former determines quark flavor transitions and the latter determines hadronic structures. 
Experimental measurements such as the determination of the  Cabibbo-Kobayashi-Maskawa matrix element $\left|V_{ud}\right|$~\cite{PhysRevD.35.934}, hyperon transition form factors~\cite{Flores-Mendieta:1998tfv} and other constants~\cite{PhysRevD.87.016002,PhysRevC.92.035206} indicate a flavor SU(3) symmetry breaking in hyperon semileptonic decays. Yet there might be another approach to probe the SU(3) symmetry breaking via searching for decays which violate the $\dsdq$ selection rule. This rule was first proposed in 1958 to explain the absence of certain hyperon decay modes in experiments~\cite{Feynman:1958ty}, and required the change in strangeness ($\Delta S$) to be equal to the change in charge ($\Delta Q$) between initial and final-state hadrons. Then it became one of the basic assumptions in Cabibbo's weak interaction theory~\cite{Cabibbo:1963yz,Cabibbo:2003cu} to propose exact SU(3) symmetry for weak hadronic currents. Therefore, any violation of this rule, which is allowed by the Standard Model in second-order weak interaction as the Feynman diagram shows in Fig.~\ref{fig:feynman}, would demonstrate the existence of weak currents belonging to higher multiplets~\cite{jamesDSDQRULE1971}.

On the experimental side, no $\dsdq$ violating decays have been observed yet. More attention has been drawn to the $\dsdq$ rule tests in neutral kaon semileptonic decays. One of the reasons is that the asymmetry between the decay rates of $K^{0}_{S,L}\to\pi^{\pm}e^{\mp}\nu$ also relates to {\it CP} and {\it CPT} invariance~\cite{Bennett:1967zz,Bernabeu:2015aga} and is used as input to the $\dsdq$ rule violation parameter. The charge-asymmetry measurement of $K^{0}_{S}\to\pi^{\pm}e^{\mp}\nu$ was recently updated by the KLOE-2 Collaboration~\cite{KLOE-2:2018yif}. As for hyperons, however, the last search for $\dsdq$ violating hyperon semileptonic decays was performed nearly 40 years ago~\cite{Zyla:2020zbs}. The current upper limit on the relative branching fraction $\Gamma(\Xi^0\to\Sigma^-e^+\nu_e)/\Gamma(\Xi^0\to\Lambda\pi^0)$ was set to be $0.9\times10^{-3}$ at the $90\%$ confidence level based on 2975 $\Xi^{0}$ events~\cite{PhysRevD.10.3545} by a fixed target experiment at the Brookhaven National Laboratory in 1974. To date, the BESIII Collaboration has collected about 10 billion $J/\psi$ events and could produce over $10^{6}$ hyperon pairs via $J/\psi$ decays~\cite{Li:2016tlt}. This allows to search for many rare and forbidden hyperon decays with a double-tag technique that was developed by the MARK-III Collaboration~\cite{MARK-III:1985hbd}. Recent BESIII results with hyperon semileptonic decays comprise the studies of the decays $\Lambda\to p\mu^{-}\bar{\nu}_{\mu}$~\cite{BESIII:2021ynj} and $\Xi^{-}\to\Xi^{0}e^{-}\bar{\nu}_{e}$~\cite{BESIII:2021emv}.

In this paper, we present a search for the $\dsdq$ violating hyperon semileptonic decay $\xisemi$ based on $(1.0087\pm0.0044)\times10^{10}$ $J/\psi$ events~\cite{numjpsi} collected with the BESIII detector at the BEPCII collider. This is the first attempt to measure the absolute branching fraction of this decay process in a collider experiment. A semi-blind procedure is performed to avoid possible bias, where about 10\% of the full dataset is used to validate the analysis strategy. The final result is then obtained with the full dataset only after the analysis strategy is fixed. Throughout this paper, including the Event Selection (Sec.~\ref{sec:evtsel}) section, charge conjugation is implied.

\section{BESIII detector and Monte Carlo simulation}
The BESIII detector~\cite{BESIII:2009fln} records symmetric $e^+e^-$ collisions provided by the BEPCII storage ring~\cite{Yu:IPAC2016-TUYA01} in the center-of-mass energy range from 2.0 to 4.95~GeV, with a peak luminosity of $1\times10^{33}$~cm$^{-2}$s$^{-1}$ achieved at $\sqrt{s} = 3.77\;\text{GeV}$. The cylindrical core of the BESIII detector covers 93\% of the full solid angle and consists of a helium-based multilayer drift chamber~(MDC), a plastic scintillator time-of-flight system~(TOF), and a CsI(Tl) electromagnetic calorimeter~(EMC), which are all enclosed in a superconducting solenoidal magnet providing a 1.0~T (0.9~T in 2012) magnetic field. The solenoid is supported by an octagonal flux-return yoke with resistive plate counter muon identification modules interleaved with steel. The charged-particle momentum resolution at $1~{\rm GeV}/c$ is $0.5\%$, and the ${\rm d}E/{\rm d}x$ resolution is $6\%$ for electrons from Bhabha scattering. The EMC measures photon energies with a resolution of $2.5\%$ ($5\%$) at $1$~GeV in the barrel (end cap) region. The time resolution in the TOF barrel region is 68~ps, while that in the end cap region is 110~ps. The end cap TOF system was upgraded in 2015 using multi-gap resistive plate chamber technology, providing a time resolution of 60~ps~\cite{etof1,*etof2,*etof3}.

Simulated data samples produced with a {\sc geant4}-based~\cite{GEANT4:2002zbu} Monte Carlo (MC) toolkit, which includes the geometric description of the BESIII detector and the detector response, are used to determine detection efficiencies and to estimate background contributions. The simulation models the beam energy spread and initial state radiation in the $e^+e^-$ annihilations with the generator {\sc kkmc}~\cite{Jadach:2000ir,*Jadach:1999vf}. An inclusive MC sample includes both the production of the $J/\psi$ resonance and the continuum processes incorporated in {\sc kkmc}. All particle decays are modeled with {\sc evtgen}~\cite{Lange:2001uf,*Ping:2008zz} using branching fractions either taken from the Particle Data Group~\cite{Zyla:2020zbs}, when available, or otherwise estimated with {\sc lundcharm}~\cite{Chen:2000tv,*Yang:2014vra}. Final state radiation from charged final state particles is incorporated using {\sc photos}~\cite{Richter-Was:1992hxq}. 
For the signal MC sample, the angular distribution measured in Ref.~\cite{BESIII:2016nix} is applied for the generation of the decay $J/\psi\to\Xi^0\bar{\Xi}^0$ followed by $\bar{\Xi}^0\to\bar{\Lambda}(\to\bar{p}\pi^+)\pi^0(\to\gamma\gamma)$ and $\Xi^0\to\Sigma^-(\to n\pi^-)e^+\nu_e$, where $\Xi^0\to\Sigma^-e^+\nu_e$ is generated with a uniform phase space model.

\section{Event selection}\label{sec:evtsel}
The $\Xi^0(\bar{\Xi}^0)$ hyperons are produced in pairs via the decay process of $J/\psi\to\Xi^0\bar{\Xi}^0$ at the center-of-mass energy of $\sqrt{s}=3.097$ GeV, and therefore can be studied with a double-tag technique. 
First, we reconstruct the $\bar{\Xi}^{0}$ candidate via the decay $\bar{\Xi}^0\to\bar{\Lambda}(\to\bar{p}\pi^+)\pi^0(\to\gamma\gamma)$. Then, we search for the signature of the signal decay $\xisemi$ from the system recoiling against the $\bar{\Xi}^{0}$. For convenience, the $\bar{\Xi}^{0}$ candidate is referred to as ``Single Tag" (ST) while the $\Xi^{0}$ candidate of signal decay is referred to as ``Double Tag" (DT). The absolute branching fraction of the signal decay is extracted by 
\begin{equation}
\mathcal{B}_{\rm{sig}}=\frac{N_{\rm{DT}}/\epsilon_{\rm{DT}}}{N_{\rm{ST}}/\epsilon_{\rm{ST}}},
\label{equ:bsigDT}
\end{equation}
where $N_{\rm ST}(N_{\rm DT})$ is the observed ST(DT) yield and $\epsilon_{\rm ST}(\epsilon_{\rm DT})$ is the corresponding detection efficiency.

\subsection{ST selection}
Charged tracks detected in the MDC are required to be within a polar angle ($\theta$) range of $|\rm{cos\theta}|<0.93$, where $\theta$ is defined with respect to the $z$-axis, which is the symmetry axis of the MDC. Photon candidates are identified using showers in the EMC. The deposited energy of each shower must be greater than 25~MeV in the barrel region ($|\!\cos\theta|< 0.80$) and greater than 50~MeV in the end cap region ($0.86 <|\!\cos\theta|< 0.92$). To exclude showers that originate from charged tracks, the angle subtended by the EMC shower and the position of the closest charged track at the EMC must be greater than 10 degrees as measured from the interaction point. To suppress electronic noise and showers unrelated to the event, the difference between the EMC time and the event start time is required to be within [0, 700]\,ns.

The $\bar{\Lambda}$ candidates are reconstructed using vertex fits from all combinations of two oppositely charged tracks, in which the one with the greater momentum is assumed to be proton and the other to be pion. The primary vertex fit constrains the tracks to originate from a common vertex, while the secondary vertex fit constraints the momentum of the reconstructed resonance to point back to the interaction point. The decay length of the $\bar{\Lambda}$ candidate has to be twice greater than its resolution. The invariant mass of $\bar{p}\pi^+$ combination is required to be within 5~${\rm MeV}/c^{2}$ from the known $\Lambda$  mass~\cite{Zyla:2020zbs} which is 3$\sigma$ of its mass resolution. The $\pi^0$ candidates are reconstructed using a kinematic fit from all combinations of two photons by constraining their invariant mass to the known $\pi^0$  mass~\cite{Zyla:2020zbs}. High-quality fit requires the $\chi^2$ value to be less than 25, and the invariant mass of two photons before the fit to be in the range of (115, 150) ${\rm MeV}/c^{2}$. Candidates with both photons from end cap EMC regions are rejected due to bad resolution. The $\bar{\Xi}^0$ candidates are reconstructed from all combinations of $\bar{\Lambda}$ and $\pi^0$ candidates described above, and the invariant mass $M_{\bar{\Lambda}\pi^{0}}$ is required to be within 20 ${\rm MeV}/c^{2}$ from the known $\bar{\Xi}^{0}$ mass~\cite{Zyla:2020zbs} corresponding to 3$\sigma$ of its invariant mass resolution. If there are multiple $\bar{\Xi}^0$ candidates that survive, the one with minimum $\left|\Delta M\right|=\left|M_{\bar{\Lambda}\pi^{0}}-m_{\bar{\Xi}^{0}}\right|$ is retained for further analysis, where $m_{\bar{\Xi}^{0}}$ refers to the known $\bar{\Xi}^{0}$ mass.

\begin{figure}[htbp]\centering
\includegraphics[width=0.9\linewidth]{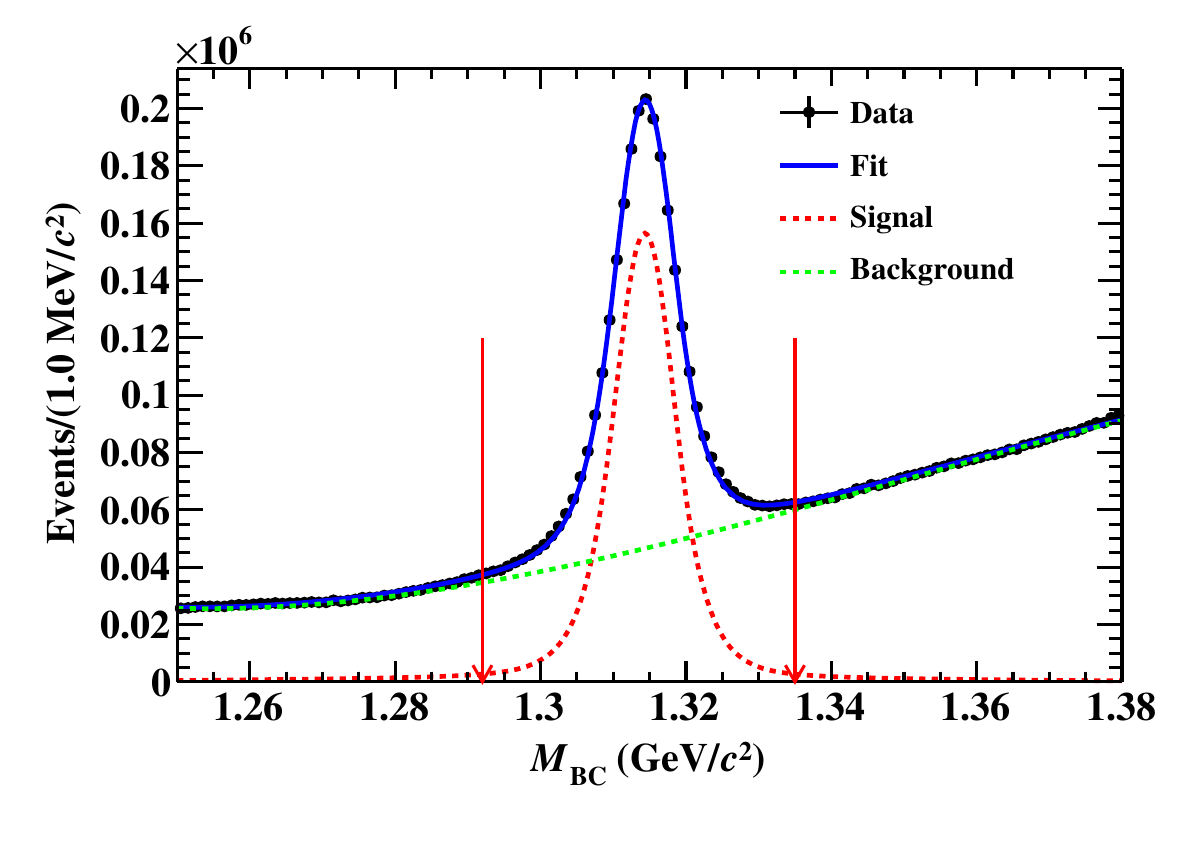}
\caption{$M_{\rm BC}$ distribution of ST $\bar{\Xi}^0$ candidates with fit results overlaid. The black dots with error bar represent data, the blue curve represents the fit result, the red dashed curve represents the signal component and the green dashed curve represents the background component. The red arrows indicate the ST signal region.}\label{fig:ST}
\end{figure}

The yield of ST $\bar{\Xi}^0$ hyperons is obtained from a binned maximum likelihood fit to the distribution of the beam-constrained mass defined as
\begin{equation}
M_{\rm BC}=\sqrt{E^{2}_{\rm beam}/c^4-\left|\Vec{p}_{\bar{\Lambda}\pi^{0}}\right|^2/c^2},
\label{equ:mbc}
\end{equation}
where $E_{\rm beam}$ is the beam energy and $\Vec{p}_{\bar{\Lambda}\pi^{0}}$ is the momentum of reconstructed $\bar{\Lambda}\pi^{0}$ combination in the center-of-mass system. In the fit, the signal shape is modeled by the MC-simulated shape convolved with a Gaussian function to account for the resolution difference between data and MC simulation. The background shape is described by a second-order Chebychev polynomial given the fact that no peaking background is observed from analyzing the inclusive MC sample. The fit result is shown in Fig.~\ref{fig:ST}. The ST yield is measured to be $1,855,681\pm1,865$ in the signal region of (1.292, 1.335) ${\rm GeV}/c^{2}$, and the corresponding efficiency is $(12.23\pm0.01)\%$ by performing the same fit procedure to the inclusive MC sample. All these uncertainties are statistical only.

\subsection{DT selection}
In the presence of ST $\bar\Xi^0$ hyperon, the signal decay $\xisemi$ is selected with the remaining tracks which have not been used in the tag side. The $\Sigma^-$ candidate is reconstructed via the $\Sigma^-\to n\pi^-$ decay. Due to the absence of a hadronic calorimeter in the BESIII detector, neutron reconstruction is challenging~\cite{BESIII:2021tbq}. We treat both the neutron and neutrino as missing particles. Only the $\pi^-$ and $e^+$ of the signal side are selected. The signature of the chosen $\pi^-e^+$ combination has been verified to sufficiently separate the signal decay from other known $\Xi^{0}$ decays. The DT signal yield is measured from the fit to the $M_{\rm BC}$ distribution of ST side because no similar observables could be constructed at the DT side.

After excluding the two charged tracks that have been used for tag reconstruction, the number of remaining tracks is required to be two, one positive and one negative charge. Particle identification~(PID) for pion is applied by combining measurements of the d$E$/d$x$ in the MDC and the flight time in the TOF to form likelihood $\mathcal{L}(h)~(h=\pi,K,p)$ for each hadron $h$ hypothesis. Tracks are identified as pions when the pion hypothesis has the greatest likelihood ($\mathcal{L}(\pi)>\mathcal{L}(K)$ and $\mathcal{L}(\pi)>\mathcal{L}(p)$) and $\mathcal{L}(\pi)$ is greater than 0.001. Positron PID uses the measured information in the MDC, TOF and EMC. The combined likelihoods ($\mathcal{L}'$) under the positron, proton, pion, and kaon hypotheses are obtained. Positron candidates are required to satisfy $\mathcal{L}'(e)>0.001$ and $\mathcal{L}'(e)/(\mathcal{L}'(e)+\mathcal{L}'(\pi)+\mathcal{L}'(K))>0.8$.

The narrow phase space leads to low momentum of the signal positron; thereby its reconstruction is challenging. Moreover, the momentum of the signal pion falls into the range where the $e/\pi$ separation ability of the PID algorithm is limited due to their similar d$E$/d$x$ responses~\cite{Asner:2008nq}. By investigating the inclusive MC sample, the dominant background components are found to be the events with $\pi^+\pi^-$ and $e^+e^-$ final states at the DT side due to high $e/\pi$ misidentification rate.

In order to reduce $e/\pi$ misidentification, we extract a variable $\chi_{\mathrm{d}E/\mathrm{d}x}$ to quantify the d$E$/d$x$ response defined as
\begin{equation}
\chi_{\mathrm{d}E/\mathrm{d}x}=\frac{\left(\mathrm{d}E/\mathrm{d}x\right)_{\mathrm{mea}}-\left(\mathrm{d}E/\mathrm{d}x\right)_{\mathrm{exp}}}{\sigma_{\mathrm{d}E/\mathrm{d}x}},
\label{equ:chidedx}
\end{equation}
where $\left(\mathrm{d}E/\mathrm{d}x\right)_{\mathrm{mea}}$ is the d$E$/d$x$ measurement for a charged track, $\left(\mathrm{d}E/\mathrm{d}x\right)_{\mathrm{exp}}$ is its expected value under certain particle hypothesis and $\sigma_{\mathrm{d}E/\mathrm{d}x}$ is its resolution. We require the $\chi_{\mathrm{d}E/\mathrm{d}x}$ value for the electron track in pion hypothesis to be less than $-4.5$, and the $\chi_{\mathrm{d}E/\mathrm{d}x}$ value for the pion track in electron hypothesis to be less than $-2.5$. In addition, the momentum of $\pi^-$ is required to be within (0.20, 0.38) ${\rm GeV}/c$ and the momentum of $e^+$ must be less than 0.20 ${\rm GeV}/c$. All the requirements except the momentum of $e^+$ are optimized using the Punzi significance~\cite{Punzi:2003bu} defined as $\epsilon/(1.5+\sqrt{B})$, where $\epsilon$ denotes the signal efficiency obtained from the signal MC sample and $B$ is the number of background events in the inclusive MC sample.

Potential large discrepancies between data and MC simulation are considered in two aspects. First, the efficiencies due to the $\chi_{{\rm d}E/{\rm d}x}$ requirement are studied with the control samples of $e^+e^-\to\gamma e^+e^-$ for the electron track and $J/\psi\to\pi^{+}\pi^{-}\pi^{0}$ for the pion track. These control samples are weighted to have the same momentum magnitude and angular distributions for the electron and pion tracks as in our signal MC sample. The ratio of the acceptance efficiencies of the $\chi_{{\rm d}E/{\rm d}x}$ requirement between data and MC simulation is $0.64\pm0.16$. Second, the $e^+e^-$ associated background is mainly produced when a photon interacts with the detector material and converts into an electron-positron pair~\cite{Asner:2008nq}. A control sample of $e^+e^-\to\gamma(\to e^+e^-)e^+e^-$ at $\sqrt{s}=3.097$ GeV is chosen to investigate the photon conversion effect. The ratio of the detection efficiencies of photon-conversion related background events between data and MC simulation is determined to be $2.929\pm0.026$. We hence correct our DT efficiency by the first factor, and consider the second factor in the estimation of the peaking background. The residual uncertainties of the two factors are taken as systematic uncertainties in Sec.~\ref{sec:syserr}. Finally, the DT efficiency is determined to be $(5.58\pm0.04)\times10^{-3}$, where the uncertainty is statistical only.

\begin{figure}[htbp]\centering
\includegraphics[width=0.9\linewidth]{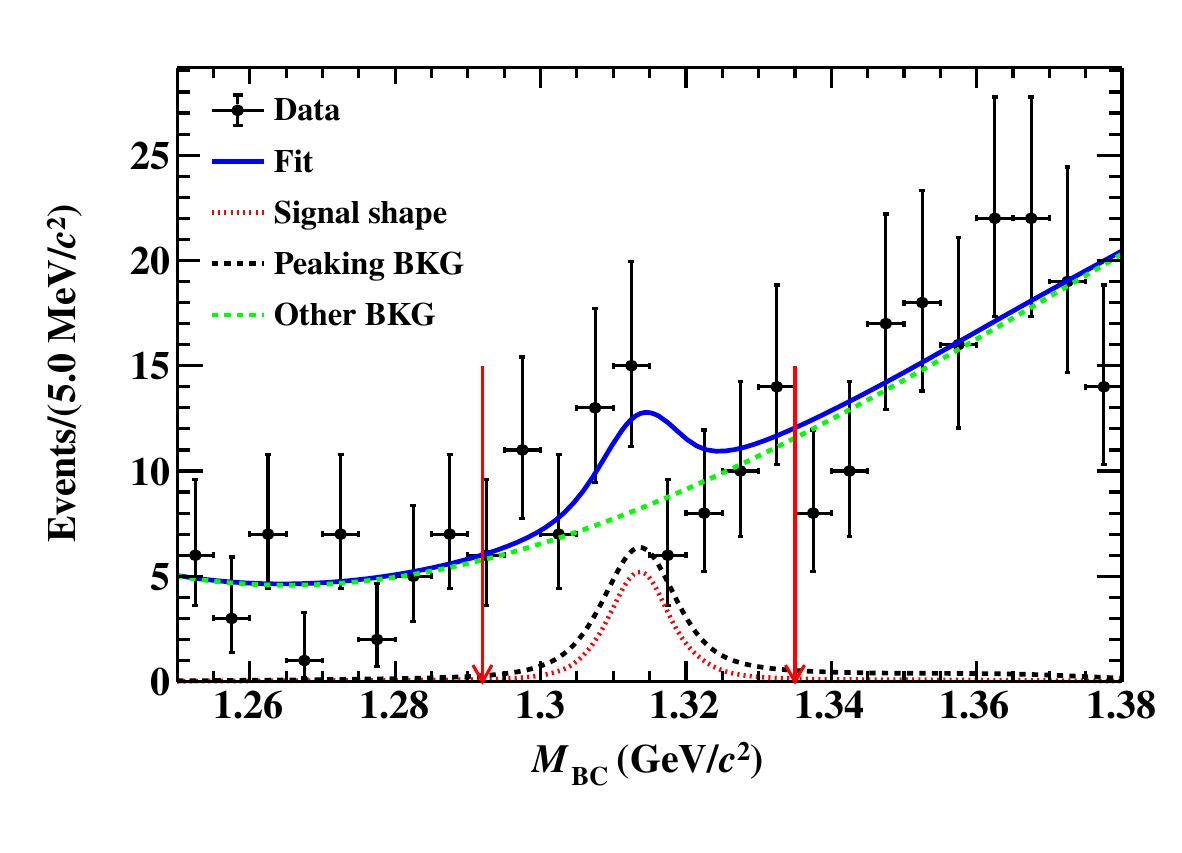}
\caption{The $M_{\rm BC}$ distribution of the ST side for DT $\Xi^0$ candidates with fit results overlaid. The black dots with error bar represent data, the blue curve represents the fit result, the black dashed curve represents the peaking background component and the green dashed curve represents the other background component. The red dotted curve shows the signal shape normalized to the branching fraction $\mathcal{B}(\xisemi)=1.6\times10^{-4}$. The red arrows indicate the DT signal region.}\label{fig:DT}
\end{figure}

The DT yield is measured by performing an unbinned maximum likelihood fit to the $M_{\rm BC}$ distribution of ST side for DT candidates. Study of the inclusive MC sample with a generic event type analysis tool, TopoAna~\cite{Zhou:2020ksj}, indicates that there is a peaking background of $J/\psi\to\Xi^{0}(\to\Lambda\pi^0)\bar{\Xi}^0(\to\bar{\Lambda}\pi^0)$ decays with a photon coming mainly from a soft $\pi^0$ and converting into a $e^+e^-$ pair in the final state. In the fit procedure, the signal shape is modeled by the signal MC simulation. The shape of peaking background is extracted from a $J/\psi\to\Xi^{0}(\to\Lambda\pi^0)\bar{\Xi}^0(\to\bar{\Lambda}\pi^0)$ MC sample. The number of peaking background events is fixed to be $23.5\pm4.1$, where the uncertainty is statistical only. It is calculated as 
\begin{equation}
N_{\rm peaking}=N_{J/\psi}\times\mathcal{B}_{J/\psi\to\Xi^0\bar{\Xi}^0}\times\epsilon_{\rm MC},\label{eqn:numpeak}
\end{equation} 
where $N_{J/\psi}$ is the number of $J/\psi$ events~\cite{numjpsi}, $\mathcal{B}_{J/\psi\to\Xi^0\bar{\Xi}^0}$ is the branching fraction cited from the Particle Data Group~\cite{Zyla:2020zbs} and $\epsilon_{\rm MC}=(1.99\pm0.34)\times 10^{-7}$ is the detection efficiency after considering the data-MC difference mentioned above. The other background shape is described by a second-order Chebychev polynomial function. Figure~\ref{fig:DT} shows the fit result where no significant signal is observed. The DT yield is extracted to be $-4.9\pm8.6$ in the signal region of (1.292, 1.335) ${\rm GeV}/c^{2}$, where the uncertainty is statistical only.

\section{Systematic uncertainty}\label{sec:syserr}
The systematic uncertainties related to the ST selection mostly cancel since we adopt the tagging technique in this analysis. The remaining systematic uncertainties involved in the branching fraction determination are classified into two types: multiplicative and additive. 

The multiplicative uncertainties, which affect the efficiency, are summarized in Table~\ref{tab:syserr} including the tracking efficiency and PID efficiency, ST fit, tag bias, number of charged tracks, selection criteria and MC model. The uncertainty due to tracking efficiency is set to be 1.0\% for the signal pion from a study of $J/\psi\to pK^{-}\bar{\Lambda}+c.c.$ and $J/\psi\to\Lambda\bar{\Lambda}$ decays~\cite{BESIII:2012xdg}, and 0.8\% for the signal electron studied with a control sample of radiative Bhabha events of $e^+e^-\to\gamma e^+e^-$. The uncertainty arising from PID efficiency is 1.0\% for the signal pion cited from a study of $J/\psi\to \pi^{+}\pi^{-}p\bar{p}$ and $J/\psi\to\pi^{+}\pi^{-}\pi^{0}$~\cite{BESIII:2018cls}, and 2.1\% for the signal electron with the same control sample used for the tracking efficiency. The uncertainty related to the ST fit is assigned to be 1.6\% by varying the signal shape description, background shape description, fit range and bin size. The tag bias effect arises from the difference of ST efficiencies obtained in inclusive and signal MC samples due to different ST reconstruction environments, and is determined to be 0.3\% following the method described in Ref.~\cite{BESIII:2021wwd}. The systematic uncertainty from the requirements on the DT charged tracks is set to be 3.2\% using the control samples of $e^+e^-\to\gamma e^+e^-$ for the electron track and $J/\psi\to\pi^{+}\pi^{-}\pi^{0}$ for the pion track. To estimate the uncertainty due to the MC model, we reweight the differential decay width distribution of the signal MC sample from the phase space model to the theoretical formula described in Ref.~\cite{Wang:2019alu}. The largest deviation of DT efficiencies before and after the reweighting, 6.8\%, is taken as the systematic uncertainty. The total multiplicative uncertainty is estimated to be 8.5\% by adding these uncertainties quadratically.

\begin{table}[!h]
\begin{center}
\small
\caption{The multiplicative systematic uncertainties.}\label{tab:syserr}
\begin{tabular}{lc} \hline \hline
Source & Uncertainty (\%) \\\hline
Tracking efficiency & 1.8\\
PID efficiency & 3.1\\
ST fit & 1.6\\
Tag bias & 0.3\\
Selection criteria & 3.2\\
MC model & 6.8\\ \hline
Total & 8.5\\ \hline \hline
\end{tabular}
\end{center}  
\end{table}

The additive systematic uncertainties mainly come from the fitted DT yield. The associated effects are examined by using alternative signal shape, peaking background yield, and other background shape. For the signal shape, we change its description from the shape directly extracted from signal MC sample to a double Gaussian function with fixed parameters obtained from fitting signal MC sample. For the peaking background, the shape is varied in the same way as for the signal shape, and the fixed yield is shifted within $\pm1\sigma$ of statistical uncertainty. For the other background shape, the alternative background shapes are chosen to be the one derived from inclusive MC sample after excluding peaking background components, and a first- or second-order Chebyshev polynomial function. Since these uncertainties are obtained with a very limited sample, they may not follow the Gaussian distribution and will be treated conservatively~\cite{BESIII:2021tfk}.

\section{Results}
\begin{figure}[htbp]\centering
\includegraphics[width=\linewidth]{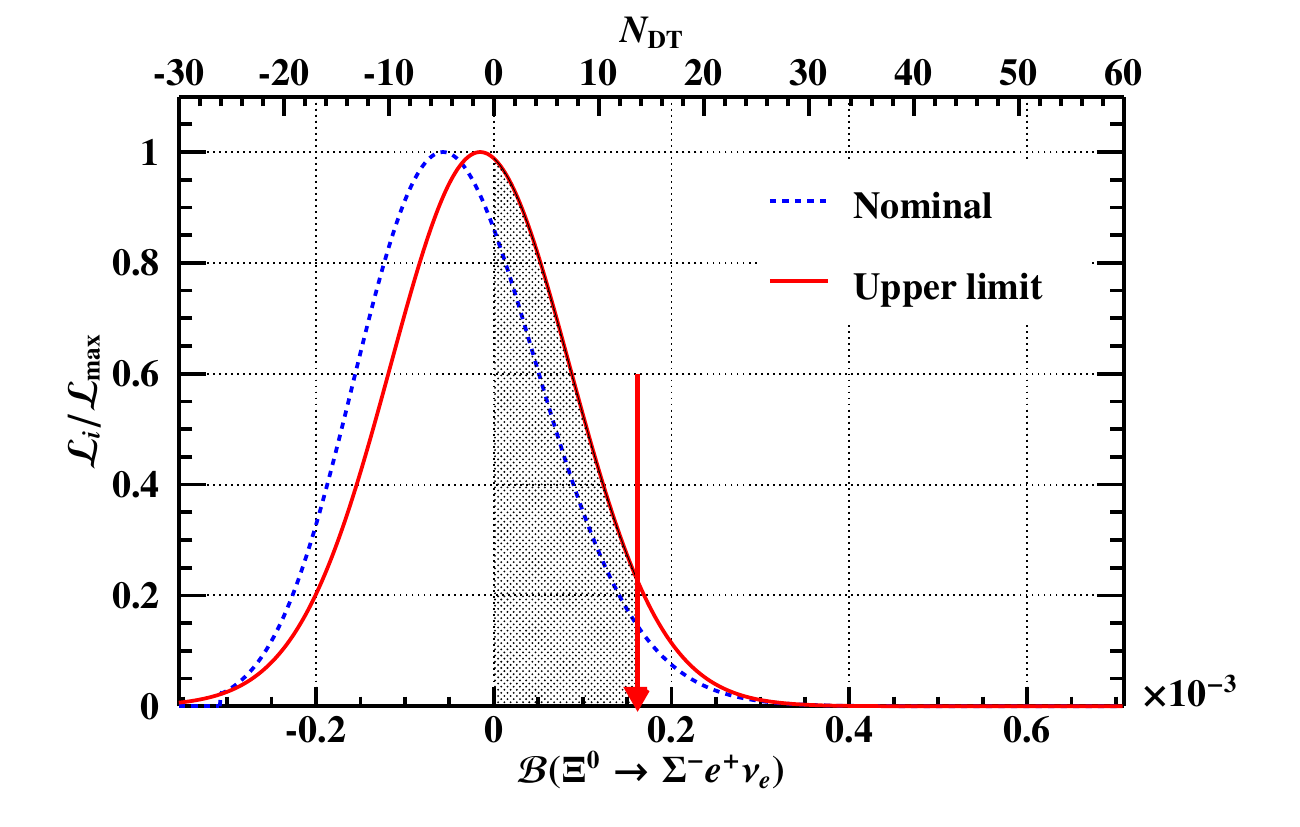}
\caption{The normalized likelihood distributions before and after incorporating systematic uncertainties. The blue dashed curve represents the raw likelihood distribution. The red solid curve represents the updated likelihood distribution after incorporating the systematic uncertainties. The top $x$-axis is for the number of signal events and the bottom $x$-axis is for the corresponding branching fraction. The shadowed area represents the integration region and the red arrow indicates the upper limits at the 90\% confidence level.}
\label{fig:finalresult}
\end{figure}

As there is no significant signal observed in data, the upper limit on the branching fraction $\mathcal{B}(\xisemi)$ is set using a Bayesian method described in Ref.~\cite{Stenson:2006gwf}. We perform a series of maximum likelihood fits to the $M_{\rm BC}$ distribution with signal yield $n$ fixed to a scan value, and obtain the corresponding maximum likelihood values to form a discrete likelihood distribution $\mathcal{L}(n)$. For the systematic uncertainties, the additive items are firstly incorporated by varying the DT fit method
, and the most conservative upper limit result is retained. Then the likelihood distribution is smeared with the multiplicative uncertainty by
\begin{equation}
\mathcal{L}'(n)\propto\int^{1}_{0}\mathcal{L}\Big(n\cdot\frac{\epsilon}{\epsilon_{0}}\Big)e^{-\frac{(\epsilon-\epsilon_{0})^{2}}{2\sigma_{\epsilon}^{2}}}\mathrm{d}\epsilon,
\label{eqn:ulsmear}
\end{equation}
as shown in Fig.~\ref{fig:finalresult}, where $\epsilon_{0}$ is the nominal DT efficiency and $\sigma_{\epsilon}$ is the multiplicative uncertainty corresponding to the efficiency value. By integrating $\mathcal{L}'(n)$ curve up to 90\% of the area in the $n>0$ region and calculating the corresponding branching fraction using Eq.~(\ref{equ:bsigDT}), the upper limit on the branching fraction of $\xisemi$ at the 90\% confidence level is set to be $1.6\times10^{-4}$.

\section{Summary}
Based on $(1.0087\pm0.0044)\times10^{10}$ $J/\psi$ events collected with the BESIII detector at the BEPCII collider, a search for $\dsdq$ violating hyperon semileptonic decay $\xisemi$ is performed. No significant signal is observed and the upper limit on its decay branching fraction is set to be $\mathcal{B}(\xisemi)<1.6\times10^{-4}$ at the 90\% confidence level. Compared with the previous experimental result~\cite{Zyla:2020zbs}, the upper limit is improved by about an order of magnitude. This search could shed light on new studies of hyperon $\dsdq$ violating decays, and other rare and forbidden hyperon decays both theoretically and experimentally.

\section*{Acknowledgement}

The BESIII Collaboration thanks the staff of BEPCII and the IHEP computing center for their strong support. This work is supported in part by National Key R\&D Program of China under Contracts Nos. 2020YFA0406400, 2020YFA0406300; National Natural Science Foundation of China (NSFC) under Contracts Nos. 11635010, 11735014, 11835012, 11935015, 11935016, 11935018, 11961141012, 12022510, 12025502, 12035009, 12035013, 12192260, 12192261, 12192262, 12192263, 12192264, 12192265; the Chinese Academy of Sciences (CAS) Large-Scale Scientific Facility Program; Joint Large-Scale Scientific Facility Funds of the NSFC and CAS under Contract No. U1832207; CAS Key Research Program of Frontier Sciences under Contract No. QYZDJ-SSW-SLH040; 100 Talents Program of CAS; INPAC and Shanghai Key Laboratory for Particle Physics and Cosmology; ERC under Contract No. 758462; European Union's Horizon 2020 research and innovation programme under Marie Sklodowska-Curie grant agreement under Contract No. 894790; German Research Foundation DFG under Contracts Nos. 443159800, Collaborative Research Center CRC 1044, GRK 2149; Istituto Nazionale di Fisica Nucleare, Italy; Ministry of Development of Turkey under Contract No. DPT2006K-120470; National Science and Technology fund; National Science Research and Innovation Fund (NSRF) via the Program Management Unit for Human Resources \& Institutional Development, Research and Innovation under Contract No. B16F640076; STFC (United Kingdom); Suranaree University of Technology (SUT), Thailand Science Research and Innovation (TSRI), and National Science Research and Innovation Fund (NSRF) under Contract No. 160355; The Royal Society, UK under Contracts Nos. DH140054, DH160214; The Swedish Research Council; U. S. Department of Energy under Contract No. DE-FG02-05ER41374.


%

\end{document}